\documentstyle[12pt]{article}

\begin{document}

\def\nn{\nonumber \\}
\def\be{\begin{equation}}
\def\ee{\end{equation}}
\def\ba{\begin{eqnarray}}
\def\ea{\end{eqnarray}}
\def\la{\label}
\def\re{(\ref}

\def\i{{\rm i}}
\let\a=\alpha \let\b=\beta \let\g=\gamma \let\d=\delta
\let\e=\varepsilon \let\z=\zeta \let\h=\eta \let\th=\theta
\let\dh=\vartheta \let\k=\kappa \let\l=\lambda \let\m=\mu
\let\n=\nu \let\x=\xi \let\p=\pi \let\r=\rho \let\s=\sigma
\let\t=\tau \let\o=\omega \let\c=\chi \let\ps=\psi
\let\ph=\varphi \let\Ph=\phi \let\PH=\Phi \let\Ps=\Psi
\let\O=\Omega \let\S=\Sigma \let\P=\Pi \let\Th=\Theta
\let\L=\Lambda \let\G=\Gamma \let\D=\Delta

\def\0{\over } \def\1{\vec } \def\2{{1\over2}} \def\4{{1\over4}}
\def\5{\bar } 
\def\6{\partial }
\def\7#1{{#1}\llap{/}}
\def\8#1{{\textstyle{#1}}} \def\9#1{{\bf {#1}}}

\def\({\left(} \def\){\right)} \def\<{\langle } \def\>{\rangle }
\def\[{\left[} \def\]{\right]} \def\lb{\left\{} \def\rb{\right\}}
\let\lra=\leftrightarrow \let\LRA=\Leftrightarrow
\let\Ra=\Rightarrow \let\ra=\rightarrow
\def\ul{\underline}

\let\ap=\approx \let\eq=\equiv  
        \let\ti=\tilde \let\bl=\biggl \let\br=\biggr
\let\bi=\choose \let\at=\atop \let\mat=\pmatrix
\def\CL{{\cal L}}


\begin{titlepage}
\renewcommand{\thefootnote}{\fnsymbol{footnote}}
\renewcommand{\baselinestretch}{1.3}
\hfill  TUW - 92 - 13 \\
\medskip
\vfill

\begin{center}
{\LARGE {Canonical Quantization of Non-Einsteinian}
 \\ \medskip  {Gravity and the Problem of Time}}
\medskip
\vfill

\renewcommand{\baselinestretch}{1} {\large {PETER
SCHALLER\footnote{e-mail: schaller@email.tuwien.ac.at} \\
\medskip THOMAS STROBL\footnote{e-mail:
tstrobl@email.tuwien.ac.at} \\ \medskip\medskip
\medskip \medskip
Institut f\"ur Theoretische Physik \\
Technische Universit\"at Wien\\
Wiedner Hauptstr. 8-10, A-1040 Vienna\\
Austria\\} }
\end{center}

\vfill
\begin{center}
submitted to: {\em Classical and Quantum Gravity}
\end{center}
\vfill
\renewcommand{\baselinestretch}{1}                          

\begin{abstract}
For a 1+1 dimensional theory of gravity with torsion different
approaches to the formulation of a quantum theory are presented.
They are shown to lead to the same finite dimensional
quantum system. Conceptual questions of quantum gravity like e.g.\ the
problem of time are discussed in this framework.
\end{abstract}

\vfill
\hfill Vienna, November 1992  \\
\end{titlepage}

\renewcommand{\baselinestretch}{1}                                       


\section{Introduction}

The canonical quantization of general relativity  is, though considerably
simplified by the introduction of the Ashtekar variables \cite{Ash} and
the loop variables \cite{Rov}, still an unsolved problem. Among the main
open issues are \cite{Les}: regularizing and
solving the quantum constraints to get (all) the physical states, finding  all
Dirac observables and  the correct inner product, interpreting the obtained
theory and, as  part of this, (re)introducing the notion of space and  time.

In \cite{1} the action for 2--dim gravity with torsion was given:
\be
{\cal L} \, =  e \, (- \frac{\gamma}{4} \, R^2 +
\frac{\beta}{2} \,T^2 - \lambda).
 \la{action}  \ee
The classical solutions of the equations of motion
were calculated in \cite{1}, \cite{2} and a
Hamiltonian formulation of the theory was provided in \cite{3a}, \cite{3b}.

Quantizing the Hamiltonian system derived from \re{action}), one
faces the same conceptual problems as in a four-dimensional
theory of quantum gravity. The calculations, however, are much
simpler.  In particular, all classical solutions of the theory
are known locally, and, similarly to $2+1$--dim gravity
\cite{Wit}, the phase space is found to be finite dimensional.
Thus we think this theory to provide a good scenario for testing
general concepts of quantum gravity. This is the motivation
behind the present paper.

Those aspects of the classical theory, which
are important for the quantization of the system are comprised in section 2.
In order to have the presentation selfcontained and to avoid
confusion about the notation, material is included in this section
which is already contained in previous
publications on the subject. But also new aspects are provided,
among them local analytic solutions around points of vanishing torsion,
which are missing in the literature.
Furthermore we show that the phase space is two dimensional if one assumes
 the space time manifold $M$ to be of the
form $M=S^1\times R$. For $\g \l <0$ and  spacelike $S^1$ the reduced
phase space has a simple topology and
a quantum theory on it can be  formulated easily (section 3.1).

In section
3.2 the Dirac method \cite{Dir} for the quantization of constrained
systems is employed:
The variables of the unconstrained phase space are quantized
in a canonical way.
The space of physical wave functions is then identified
with the kernel of the
quantized constraints. Finally
an inner product is to be introduced in the space of physical wave functions.
We present two ways to achieve that: One proposed in  \cite{Hen}
starts from a measure in the unconstrained phase space, which is reduced by
gauge conditions. Since the Faddeev-Popov
determinant turns out to be inadequate to guarantee gauge independence in our
case, the method is altered somewhat.
No gauge conditions are needed for constraints which act
multiplicatively. The gauge conditions implicitely introduce an internal
time into the system.
A somewhat different approach \cite{Les} defines the
inner product without any reference to a measure in the
unconstrained phase space by requiring a sufficiently large set
of Dirac observables to be hermitian.
We conclude the subsection by applying also the simple quantization scheme
used in \cite{Wit}
to quantize $2+1$ dim gravity with zero cosmological constant.

Under the restriction $M=S^1\times R$ with spacelike $S^1$ the
quantum theories resulting from the reduced phase space
quantization and the Dirac method are shown to be equivalent.
In the Dirac approach, however, one need not know the topology
of the reduced phase space. The method is still applicable, if
the above restrictions are lessened.

An elegant alternative is presented in section 3.3:
The constraints are abelianized before quantization. It is most remarkable
that this is possible in a relatively simple way. Exploiting the fact
that the abelianized constraints may serve as canonical variables,
the quantization becomes simple now. The resulting quantum system
is completely equivalent to the one obtained in the preceding sections.

Physical questions usually refer to space--time events characterized
by coordinates $x^\mu$.
To answer them in a quantum theory of gravity
they have to be reformulated in terms of Dirac
observables --- the space time coordinates $x^\m$ enter as
parameters. As in the classical theory the choice of a
coordinate system is equivalent to the choice of a gauge.
This idea is realized in section 4:
A one parameter family of gauge conditions allows to express
gauge dependent quantities in terms of the Dirac observables.
Once the parameter in this family is interpreted as an intrinsic
time, these quantities become dynamical (i.e.\ time dependent).
With respect to this dynamics, a Schroedinger picture is formulated.
The gauge independent formulation of the quantum states then
corresponds to the according Heisenberg picture.
  In either picture it is possible
to predict quantities such as $\< g_{\m\n} (x^\m) \>$ or
$\< T^2 (x^\m) \>$.


\section{The Classical Theory}

In the Lorentz-bundle over the space-time manifold we will use the metric
       \begin{equation} \eta_{ab} =\left( \begin{array}{cc}  0 &  1 \\
1 & 0 \end{array}
\right),
\label{lormet} \end{equation}
with   $a, b \in \{ +,-\}$. In \re{action}) $R$ is the Ricci
scalar and $T^a$ is the Hodge dual of the torsion two--form
($T^2 \equiv T^aT_a$).  Thus they are (Lorentz vector valued)
functions on the space time manifold. A simple rescaling allows
us to   set $\b = 1$ in the following.

The calculation of the canonical conjugate momenta to the
components $e_{\mu}{}^{\pm}$ and $\omega_{\mu}$, $\mu \in \{0,1\}$, of the
zweibein and the connection yields  the
relations ($\dot \ph = \6 \ph / \6 x^0, \6 \ph = \6 \ph / \6 x^1$)
$$        \pi_a := {\partial {\cal L} \over \partial \dot{e}_1{}^a} \,= \,-
                       T_a     $$
$$        \pi_{\omega} := {\partial {\cal L} \over \partial \dot{\omega}_1}
   \,= \,           \gamma \, R  $$
and the primary constraints
           \be {\partial {\cal L} \over \partial \dot{e}_0{}^a} =
              {\partial {\cal L} \over \partial \dot{\omega}_0{}^a} =0.
           \la{primary} \ee
It is remarkable that torsion and curvature serve as canonical momenta
for the 1-components of zweibein and connection. The canonical Hamiltonian
is
     \be H = - \int e_0{}^a \, G_a + \omega_0 \, G_\omega \la{Ham} \ee
with the secondary   constraints ($\varepsilon_{+-}=1$)
\addtocounter{equation}{1}
$$ G_a \,=\, \varepsilon_{ab} \,  E \, e_1{}^b - \varepsilon^b{}_a  \, \pi_b \,
\omega_1 + \partial \pi_a \; \approx \; 0 \eqno(\arabic{equation}a) $$
$$ G_{\omega} \,=\, \varepsilon^a{}_b \, \pi_a \, e_1{}^b
+  \partial \, \pi_{\omega} \;
\approx \; 0, \eqno(\arabic{equation}b) $$
in which we used the abbreviations
\begin{equation}
 E \; \equiv \; \frac{1}{4\gamma} \, (\pi_{\omega})^2
  - \frac{1}{2} \, \pi^2 - \lambda, \quad \pi^2 \, \equiv
   \, \pi^a \, \pi_a.
 \la{E}
\end{equation}

Anticipating our later restriction to $x^1 \in S^1$, we have omitted
a surface term in \re{Ham}) resp. (5).
Because of \re{primary}) $e_0{}^a$ and $\o_0$ act as mere
Lagrange multipliers within the Hamiltonian \re{Ham}). They resemble
lapse and shift in 4d gravity. The constraints (5) are
first class:
\addtocounter{equation}{1}
$$ \{ G_a, G_{\omega} \}\, =\, -  \varepsilon^b{}_a \, G_b \, \delta
\eqno(\arabic{equation}a) $$
$$ \{ G_a, G_b \} \,=\, \varepsilon_{ab} \,
( -\pi^c \, G_c +
\frac{\textstyle 1}{\textstyle 2\gamma}
\, \pi_{\omega} \, G_{\omega} ) \, \delta .
\eqno(\arabic{equation}b) $$

Let us analyze briefly the geometrical meaning of the constraints.
$G_\omega (x^1)$ generates local Lorentz
transformations, thus corresponding exactly to the Gauss
constraint in the Ashtekar formulation of general relativity: The first term of
(5b), $e_1{}^-
\, \pi_- - e_1{}^+\, \pi_+$, is the  generator of
O(1,1)--transformations in the 4 dim. phasespace ($e_1{}^a$,
 $\pi_a$), the second term reflects  the
transformation property of the connection $\o$ under local
Lorentz transformations on a  slice of the space
time manifold with $x^0 = const$.
The combination
\be e_1{}^a\, G_a + \omega_1 \,
G_\omega = e_1{}^a\, \partial \pi_a + \omega_1 \, \partial
\pi_\omega \la{diff} \ee
of the constraints can easily be shown to generate
diffeomorphisms $\delta x^1 = \epsilon(x^1)$ [cf.\ also \re{Ham})],
thus being the analogue of the
vector constraint of the usual $3+1$--theory. The combination
\be\pi^a \, G_a - E\,G_\omega = \6 (\pi_+\pi_-) -E \6 \pi_\o = {1\0 4\g}
\exp(-\pi_\omega)  \, \partial Q \la{dQ} \ee
with
\be Q := \exp (\pi_\omega)
\, [2\g \, \pi^2 -
(\pi_\omega - 1)^2 -1 +\Lambda ], \quad \Lambda \equiv 4\g\l
\la{Q}  \ee
generates, up to local Lorentz transformations,
diffeomorphisms in the direction of constant curvature and
torsion squared, since it is  a polynomial in the momenta only.
The quantity Q defined in \re{Q}) has vanishing Poisson brackets
with all the constraints, and therefore it is a constant on any
classical solution of the field equations  \cite{2},
\cite{3a}.  As a consequence of this and
\re{Q}) lines of constant curvature always coincide with lines
of constant $T^2$.

Another combination of the constraints, generating diffeomorphisms
in the $x^0$ direction, is the  Hamiltonian \re{Ham}).
The gauge choice $e_0{}^+=1, \, e_0{}^-=0, \, \omega_0=0$ in \re{Ham})
(light cone gauge) identifies it with
$-G_+$. As this choice
corresponds to a space time metric with $g_{00} =0$, it is obvious
that $G_+$ generates diffeomorphisms in a lightlike direction (up to
local Lorentz transformations, again). A similar
argument holds for $G_-$. To complete the analogy to the $3+1$ dim. theory,
one can choose an appropriate linear combination of $G_+$ and $G_-$ to
play the role of the scalar constraint.

To find the local behaviour of a solution in the neighbourhood
of a point $P$ on the space time manifold where $T_+(P) = -
\pi_+(x^\mu(P)) \neq 0$, we can satisfy the constraints by inverting them
algebraically to express $\pi_-$, $\omega_1$, and $e_1{}^+$ in terms of
$\pi_\o, \, \pi_+$, and $e_1{}^-$, as well as
$Q(x^1) = Q_0 = const$
[because of \re{dQ}) using \re{Q}) instead of $G_-$].
The general solution in a neighbourhood of P can then be determined by
applying $-G_+$ to the
remaining three fields, generating the $x^0$ dependence of the
solution
in the light cone  gauge. Thus integrating $\{\pi_+, \int \! G_+\}=0$,
$\{\pi_\omega,\int \! G_+\}=-\pi_+$, and $\{e_1{}^-,\int \!  G_+\}= -
e_1{}^- \,\pi_+$, we  end up with
\ba
\pi_{\o}& =&  \pi_+ \, x^0 + B(x^1) \nn
\pi_+ &= & A(x^1) \nn
\pi_- &=& {1\04\g\pi_+}[Q_0 \, \exp(-\pi_{\o}) + (\pi_{\o}-1)^2 +1 -\L] \nn
\omega_1 &=&  -{1\0\pi_+} \, (e_1{}^- \, E + \6 \pi_+) \nn
e_1{}^+ &=& {1\0\pi_+} \, (e_1{}^- \, \pi_- + \6 \pi_{\o}) \nn
e_1{}^- &=& D(x^1) \, \exp(\pi_{\o}), \la{l1} \ea
in which  $A, \,B, \, D$ [$A(x^1(P)) \neq 0$] are still arbitrary functions.
Nevertheless, locally it   is possible to gauge them away:
Not changing
the values of $e_0{}^a$ and $\o_0$, it is possible to obtain $A(x^1) =1$
by the choice of an appropriate Lorentz
frame,  $B(x^1) =0$ by an $x^1$--dependent shift
of the origin of the $x^0$-variable, and   $D(x^1)=1$ by an
$x^0$ independent transformation of the $x^1$ variable. This 'normalized
solution'  may be taken also to represent a solution in the
neighbourhood  of
a point $P$ with $\pi_+(P)=0, \pi_-(P)\neq 0$ since we still
have not made use of the discrete Lorentz transformation which
exchanges  '+' and '-'.  Thus {\em locally} the space of
solutions to the field eqs.\ is parametrized completely by the
value of $Q_0$ \cite{3a}, \cite{Katneu}.

The solution in the neighbourhood of a point with $\pi_+(P)=\pi_-(P)=0$
can be obtained by first calculating the flow of $G_-$ starting
from $P$ (we label this null line $x^0 = 0$):
\addtocounter{equation}{1}
$$ \pi_+(x^1) = E_0 \, x^1, \,
\pi_-(x^1) = 0,  \eqno(\arabic{equation}a) $$                       
$\pi_{\o}(x^1) = B(x^1(P)) = const$ and $E(x^1) = E_0 = const$ being
determined implicitely by $Q_0$ [cf.\ \re{Q}) and \re{E})]. For $E_0 \neq 0$
the Gauss
constraint then yields
$$ e_1{}^+(x^1)=0,   \eqno(\arabic{equation}b) $$
whereas locally a Lorentz gauge representative of the solutions to (5a) is
$$ \o_1 (x^1) = -E_0  , \, e_1{}^-(x^1) =  E_0  x^1 -1.
 \eqno(\arabic{equation}c) $$
To avoid $e = 0$ at $x^1 = 1/E_0$
one can also 'deform' $\pi_+(x^1)$ into $E_0 \, \arctan x^1$, when
changing (12c) correspondingly. Finally acting with  $-G_+$ on (12)
gives the (local) $x^0$--dependence.

Note that according to \re{Q}) such solutions (and
all solutions with points where $\pi^2\equiv 2\pi_+\pi_-=0$)
are possible only for a certain
range of $Q_0$--values. An analysis of \re{Q}) shows the existence of
a function $h(\L)$ with the following property: $\pi^2$ has zeros on $M$ for
$Q_0 < h(\L)$ and does not for $Q_0 > h(\L)$; one finds $ h$  to be zero for
$\L \le 1$ and to increase monotonically for $\L > 1$.

A solution with constant curvature and vanishing torsion (de Sitter or
Liouville solution, $\pi_\o = \pm \sqrt{\L}$)  is included in (12);
it is obtained by choosing
\be Q_0 = Q_{deS} :=2 \exp(\pm
\sqrt{\L }) \, ( \mp \sqrt{\L} -1) \; \LRA \;
E_0 = 0. \la{deS} \ee
In this case the eqs.\ (12b), (12c) are not a consequence of the constraints,
which are trivially fulfilled, but they represent a (local) gauge choice
along the line $x^0 =0$. Note that \re{l1}) gives a different solution
for $Q_0 = Q_{deS}$.

For the case $Q_0 \neq Q_{deS}$ (and $Q_0$
within the range of the existence
of $T_a=0$) one may construct the
solutions (12) from \re{l1}), if one allows
for zeros of $A(x^1)$ (but $\6 A (P) \neq 0$)
and chooses $B$ and $D$ such that singularities are avoided. Since
in the coordinate system chosen above the curvature has the form
$\pi_\o = - E_0 \, x^1x^0 + B_0(Q_0)$, points with vanishing $T_a$
are saddle points of the curvature (and vice versa). Local solutions
around such points have not been obtained in the previous literature;
nevertheless, within the conformal gauge saddle points of $R$ appeared by
gluing together solutions \cite{Katneu}.

To do quantum mechanics purely local considerations are not
sufficient. A complete analysis, however, treating all the
topological aspects of the theory is beyond  the
scope of the present paper. Instead we will restrict our considerations
to the case that  the space time manifold $M$ can be written as
$M=S^1\times R$, the $S^1$ being spacelike. Further we will
regard the case   $\L < 0$  so as to exclude $Q_0 = Q_{deS}$.

Let us prove that under these assumptions   there are no
solutions to the field eqs.\ which allow for a zero of $T_+$ or $T_-$.
The cylinder we are to consider may be covered by one chart with periodic
boundary conditions in the $x^1$ direction.
Now, having e.g.\ a point P with $\pi_-(P) = 0$, the constraint equation
(5a) as well as \re{deS}) and the  existence of a
spacelike $S^1$ through P
 tell us that
\be \6 \pi_- (P) = E_0 \, e_1{}^+ (P) \neq 0.  \la{no} \ee
Since $\pi_-$ is a periodic function in $x^1$ there has to exist at least
another point P' on this line $x^0 = const$ with vanishing $\pi_-$ and
${\rm sgn}\, \6 \pi_- (P') \neq {\rm sgn} \,
\6 \pi_- (P)$. But then eq.\ \re{no})
for P and P' implies that the sign of $ e_1{}^+$ changes along the curve
$x^0 = const$; this contradicts the assumption that it is spacelike.
Let us note in parenthesis that when regarding $M= \Sigma \times R^1$ with
$\Sigma = R^1$ spacelike, the above argumentation shows us only that there
cannot exist more than one (null) line with $\pi_- =0$ for $Q_0 \neq Q_{deS}$;
the same is true for  $\pi_+ =0$. For $Q_0 = Q_{deS}$
the above reasoning does not go through as \re{no}) is not true.

Thus in our topological setting we know  all the classical solutions.
They are given by \re{l1}) with $A(x^1) \neq 0$ and $Q_0 \ge 0$; the latter
restriction comes from the requirement
 $\pi_- \neq 0$ (cf.\ the paragraph following eq.\ (12)). But how
can we  possibly do quantum mechanics, if there is only one parameter $Q_0$
labelling the gauge inequivalent solutions? Actually for the case of
our topology there is a
second one as a simple consideration shows: For a particular
fixed value $R_0$ of $R$ the metric on the space time manifold
induces a metric on the curve generated by $R=R_0$. As this
curve is compact (cf.\ the first eq.\ of \re{l1})),
it has a finite length.  This length is
(though $R_0$ dependent) by construction gauge-invariant and
obviously  not determined by $Q_0$. (We can e.g.\ change the interval
of periodicity at will without changing the integrand, which
can be made $x^1$--independent). Thus there is a  quantity
$P_0>0$ characteristic  for the 'size of the universe'.

This fact and that
there are no further gauge invariant quantities can be seen also from a
more formal point of view: As before it  is always possible
to find a gauge such that $A(x^1)=1$ and $B(x^1)=0$. But now, normalizing the
interval of periodicity of $x^1$ to $[0,1]$,
a diffeomorphism $x^1 \ra f(x^1)$, $D(x^1) \ra \6 f \, D(f(x^1))$
cannot change the zero mode of the arbitrary (periodic)
function $D(x^1)$; therefore it is
possible only to make $D$ constant: $D(x^1) = D_0 =: - P_0 / 4\g $. The
identification $P_0 \lra - P_0$ then is obtained by the gauge transformation
$x^1 \ra - x^1$; and $P_0 \neq 0$ since we required   the
$S^1$ to be spacelike.
Thus in our topological setting the space of solutions of
the eq.o.m. (and thus the reduced phase space of the theory) is a two
parameter space:
$$  \pi_{\o} =   x^0, \qquad \pi_+ = 1  $$
$$ \pi_- =  {1\0 4\g}[Q_0 \, \exp(-x^0) + (x^0-1)^2 +1 -\L] $$
$$ \omega_1 = - e_1{}^- \, E, \qquad e_1{}^+ = e_1{}^- \, \pi_-  $$
\be e_1{}^- = -4\g \,P_0\, \exp(x^0), \la{l3} \ee
with
\be Q_0 \ge 0, \qquad  P_0 >0. \la{>} \ee
In this gauge the quantity $E$, defined in \re{E}), becomes:
\be  E =  \frac{1}{4\gamma} \, (x^0)^2
  -  \pi_- - \lambda = {1\04\g} [-Q_0 \, \exp(-x^0) +2 (x^0 -1)]. \la{Et} \ee
Because of $g_{11} = 2(e_1{}^-)^2\pi_-$
the requirement '$S^1$ be
spacelike' is compatible only with $\g >0$, whereas
 for the case $\g <0$ there exist
no such solutions to the field equations. Requiring $M=S^1 \times
R^1$, the $S^1$  being
timelike, on the other hand, one obtains \re{l3}) for $\g<0$ and  no
 solutions for  $\g >0$ ($\L<0$). This result holds
irrespective of any gauge as is obvious from
  (5b) multiplied by
 $e_1{}^-/\pi_+$
 and the fact that  $\6 \pi_\o$ cannot be definite on a circle.
As will be shown in sec.\ 4,  furthermore,  the evolution parameter $x^0$
in \re{l3}) can be taken to be purely timelike.

For the case  $\L \ge 0$ the requirement of the existence of
a spacelike closed section leads to  \re{l3}) with $Q_0 = Q_{deS}$ or
 $Q_0 > h(\L)$,
 $h$  having been defined in the
paragraph following eq.\ (12), as well as to the Liouville solution.
A gauge representative of the latter
depends also on {\em one} (topological) quantity $P_0$,
as can be shown by  considerations similar to the ones leading to
\re{l3}).
Although we do know all the classical solutions in this extended
situation, too, the construction of
a  consistent quantum theory on {\em these} classical
solutions seems hardly manageable due to the existence of a discrete part
in the spectrum of $Q_0$ (cf. sec.\ 3.1 below) --- except when assigning these
points the measure zero, certainly.
Thus for $\L \ge 0$ the classical requirement that
the $S^1$ be {\em spacelike} cannot be maintained within a quantum theory.

\section{The Quantum Theory}

For the present model the simplest and most straightforward quantization
is the reduced phase space quantization (sec.\ 3.1). It makes use of the
fact that we know all the classical gauge inequivalent solutions (under
the assumptions made in the preceding section). Nevertheless, in order to
gain insight into theories where not  all the classical solutions are known,
such as general relativity, it is instructive to apply also other
standard methods
of quantization like for instance the Dirac procedure (sec.\ 3.2).
Canonical transformations,
obtainable also without knowing the classical solutions,
can dramatically simplify the  task of quantization. This shall be
 illustrated in sec.\ 3.3, where we succeed in describing the
constraint surface by the vanishing of canonical coordinates.

\subsection{Reduced Phase Space Quantization}

Having  the reduced phase space at our disposal,
which is the quarter
of a plane under our assumptions [cf.\ \re{>})],
we need to find the symplectic structure it
inherits from the  unconstrained phase space. This can be
achieved by first finding  the Dirac observables which correspond to
$Q_0$ and $P_0$ in \re{l3}) and then by calculating their Poisson bracket.

$Q_0$  is obviously just the constant mode of \re{Q}), i.e. (as we
fixed the length  of periodicity to one,  factors $2\pi$ are avoided)
\addtocounter{equation}{1}
$$ Q_0 =  \int_{S^1} \, Q. \eqno(\arabic{equation}a) \la{P0} $$ 
To find the
gauge independent quantity corresponding to $P_0$,
 one first makes the last eq.\ of \re{l3}) exlicit in $P_0$.
Deviding the obtained expression by $\pi_+$,
which is one in the gauge of \re{l3}), it becomes    Lorentz invariant.
Integration of the resulting one--form yields the
 diffeomorphism invariant quantity
$$  P_0= -{1 \0 4 \g}\int_{S^1} \exp(-\pi_{\o}) \, { e_1{}^-\over \pi_+}
\eqno(\arabic{equation}b) $$                                             
as our second Dirac observable, commuting (weakly)
with all the constraints.
Now it is straightforward to verify that the symplectic form on the reduced
phase space equals
\be{\rm d}Q_0
\, \wedge {\rm d}P_0. \la{sym} \ee
However, $P_0$ in (18b) is   not invariant against the
discrete transformation $x^1 \ra -x^1$,  which is not included within the
(continuous) flow of the constraints. Thus the completely gauge
independent quantity corresponding to the normalized solution
\re{l3}) subject to the restriction \re{>}) is actually
\be \mid \! P_0 \! \mid \, = \sqrt{P_0^* P_0}.  \ee

Now the quantization is quite simple. The commutation relations \re{sym}),
or better the corresponding
Weyl algebra, as well as  the first eq.\ \re{>}) as a restriction to the
spectrum of $\hat Q_0$ yield an
${\cal L}^2(R_+)$ with Lebesgue measure as our Hilbert space
(cf.\ e.g.\ \cite{Thi}). In this $\hat Q_0$ acts as
a multiplicative operator and
$\hat P_0$ as the usual derivative operator (up to unitary equivalence).
And since it is $\sqrt{\hat P_0^* \hat P_0}$ which  corresponds to the
classical quantity $P_0$ in \re{l3}), we also have no problem
with self--adjointness and the second restriction \re{>})  (as we
would have with  $\hat P_0$).

Note that there is  so far no 'dynamics'
present in this formulation of the quantum theory. As typical
for theories formulated in a reparametrization invariant way our
Hamiltonian \re{Ham}) vanishes so that a priori there is no (naive)
Schroedinger eq.\ or also  no (naive) path integral. How and in how far
we can
introduce some notion of time into the canonical framework above
shall be discussed in sec.\ 4. The corresponding problem in the
path integral formulation shall be tackled elsewhere \cite{Ku}.

\subsection{Dirac quantization}

In this section we shall quantize the unconstrained phase space and then
calculate physical wave functions as the kernel of the quantized constraints.
The form of the primary constraints \re{primary}) allows to simply
eliminate the   zero components of our fields.  So we are
left with a phase space $\G$ spanned by the variables
$\o_1,\,e_1{}^a,\,\pi_{\o},\,\pi_a$.  Since our constraints (5)
are linear in the coordinates but quadratic in the momenta, we
will work in the momentum representation:
\be
\o_1 \ra i \hbar {\delta\over\delta\pi_\o}, \qquad
e_1{}^a \ra i \hbar {\delta\over\delta\pi_a}. \la{Op}
\ee
Thus the  quantum constraints become
\be
\hat G_\omega \Psi= \hat G_+ \Psi = \hat G_- \Psi =0  \la{cons}
\ee
with ($[ \; , \,]_+$ denotes the anticommutator)
\addtocounter{equation}{1}
$$ \hat G_\omega = \6\pi_\o + i\hbar(\pi_- {\d\0\d\pi_-} - \pi_+ {\d\0\d\pi_+})
 \eqno(\arabic{equation}a)  $$                                 
$$ \hat G_+ =  \6\pi_+ + i\hbar({1\0 2}[E , {\d\0\d\pi_-}]_+
+ \pi_+ {\d\0\d\pi_\o})
 \eqno(\arabic{equation}b)  $$                                 
$$ \hat G_- =  \6\pi_- - i\hbar({1\0 2}[E ,{\d\0\d\pi_+}]_+
+ \pi_- {\d\0\d\pi_\o}).
 \eqno(\arabic{equation}c)  $$                                 
As already proven in \cite{3b} the quantized constraints (23) form
a closed algebra. This is crucial for the consistency of the simple
quantization scheme used here.  Otherwise we would rely on more elaborate
techniques like e.g.\ BRST quantization (cf.\ also \cite{Jap} and sec.\ 5).
Note that the first replacement in \re{Op}) breaks
the local Lorentz covariance
present in the  classical theory: whereas  the lefthand side of that
eq.\ transforms  as  usual for a connection  of the  Lorentz group
($\pi_\pm \ra \exp[\pm \a (x)] \pi_\pm \Ra \o_\mu \ra \o_\mu -
\6_\mu \, \a$), the righthand side remains unchanged. This is a
feature which should prevail also in the Ashtekar formulation of the
$3+1$ theory.

Up to purely multiplicative terms our quantum constraints contain only
 Lie derivatives.
Thus the calculation
of the kernel of the constraint operators will simplify considerably,
if, instead of some of the momenta, we use other variables
which commute strongly with the classical constraints. Because of
\re{>}) our wave functions have their support only in an area
where $\pi_+$ and $\pi_-$ are different from zero [cf.\ \re{Q}) and remember
$\L <0$]. In such an area  the map from  either $\pi_-$ or $\pi_+$
to $Q$ is bijective. Therefore to start with we will write our wave functions
as
\be \Psi = \Psi[Q(\pi^2,\pi_\o), \pi_+, \pi_\o].  \la{ans} \ee
With this general ansatz the integration of the first two eqs.\ \re{cons})
is straightforward, yielding
\be \Psi = \exp(-{i\0 \hbar} \int_{S^1} \6 \pi_\o \ln |\pi_+|) \,
\exp({1\0 2}\d(0) \! \int_{S^1} \! \pi_\o) \; \ti \Psi [Q],
\la{wav1} \ee
whereas the last eq.\ \re{cons}) becomes
\be  \partial Q \, \ti \Psi[Q]=0, \la{par} \ee
as is also clear from \re{dQ}) which is valid also in the quantum case.
The $\delta (0)$ is understood to be defined in an appropriate
regularization.  We could e.g.\ discretize the $x^1$ variable
according to $x_i-x_{i-1}=l$. $\delta (0)$ appears to be
${1\over l}$ in this regularization.

The operator ordering in (23) guarantees that the
quantum constraints are hermitian with respect to the Lebesgue measure $\int
[{\rm d}\pi_\o][{\rm d}\pi_a]$.
We could avoid the $\delta(0)$ term by a different
choice of the operator
ordering in \re{cons}): Putting all derivatives to the right, the
constraint algebra still closes and as the constraints vanish on
physical states, they are automatically hermitian in the
physical sector, whatever operator ordering we choose.
We will find, however, that the $\delta(0)$
term plays quite a crucial role in the reduction of the Lebesgue
measure to an inner product in the space of physical states.

Starting from \re{ans}) with  $'+'$ and $'-'$ exchanged, we
obtain analogously
\be \Psi = \exp({i\0 \hbar} \int \6 \pi_\o \ln |\pi_-|)\,
\exp({1\0 2}\d(0) \! \int \! \pi_\o) \; \ti \Psi [Q]
\la{wav2} \ee
as well as \re{par}). Due to the latter eq.,
which is equivalent to setting \re{Q}) equal to some
constant $Q_0$, it is obvious that the transition amplitude
\be \exp({i\0 \hbar} \int_{S^1} \6 \pi_\o \ln |\pi^2| \,
{\rm d}x^1) = 1 \la{trans} \ee
so that the $\ti \Psi [Q]$ in \re{wav1}) and \re{wav2}) do indeed coincide.

Note that in the above considerations we made use of our restrictions on
$\L$ and the topology only when we restricted the support of the physical wave
functions to positive values of \re{Q}). Within the Dirac quantization
everything else is  the same also in the completely general case:
\re{wav1}) and \re{wav2}) fulfilling \re{par})
give the general solution to \re{cons}) on  charts of the phase space with
$\pi_+ \neq 0$ and $\pi_- \neq 0$, respectively.  Because of \re{trans}) they
can be patched together to give the wave  functions fulfilling  \re{cons}) on
all of the phase space except for points with  simultaneous zeros of
$\pi_+$ and $\pi_-$.  To extend this solution to all of the phase space except
for points with $\pi_a = E = 0$ ($\LRA Q = Q_{deS}$) by a further ansatz
$\Psi = \Psi [Q,\pi_a]$ does not seem to be so easy though: A calculation
analoguosly to the above ones yield a difficult differential eq.\ of first
order, and to make a  good guess is aggrevated by the fact that the expected
phase factor will definitely be not locally Lorentz covariant due to \re{Op})
or (23a),
as can be seen explicitely from \re{wav1}) or \re{wav2}).

However,
our solutions \re{wav1}) {\em or} \re{wav2})  extend to a much more
general situation anyway: The phase factors in  $\Psi$ can be integrated
as long as the torsion does not vanish on an {\em interval} of the $S^1$.
Thus with \re{wav1}) we exclude only functional  distributions solving
the quantum constraints \re{cons}) such as
\be
\d [\pi_a] \, \d [\pi_\o \mp \sqrt{\L}], \la{Lio}
\ee
which obviously corresponds to the Liouville or de Sitter solution.

Still we have to define an inner product in the space of wave functions
\re{wav1}).
To this end we may first realize that
$\Psi^*\Psi$ gives a factor $\prod_{x^1} \exp(\pi_\omega(x^1))$ and
that the product of this factor and the formal Lebesgue measure
$[{\rm d}\pi_\o][{\rm d}\pi_a]$  on the unconstrained momentum space yields
an expression being invariant under the classical flow of the constraints.
The integral of $\Psi^*\Psi$ with the Lesbegue measure, however,
will diverge, as the wave functions are roughly speaking constant
in the direction of $G_+$ and $G_\o$. Note that having implemented these two
constraints $G_-$ is purely multiplicative and is of no relevance for the
considerations at this stage. $G_+$ and $G_\o$ are the generators of a
non--abelian {\em group} [cf.\ (7a)], the (infinite) volume of which
has to be 'devided out'  from
the integral.
As this group acts freely and transitively on the
($\pi_\o$, $\pi_+$)--plane (or more strictly speaking the half plane
with positive values of $\pi_+$), it is suggestive to restrict
the values of $\pi_\o$ and $\pi_+$ by the gauge conditions
\be \pi_+(x^1)=c(x^1) \qquad \pi_\omega (x^1) =t(x^1).  \la{gauge1} \ee
These gauge conditions may be realized by the introduction of
$\delta[\pi_+ -c]\, \delta[\pi_\o -t]$ into the measure.
This expression is, however, not invariant under the flow of
$G_+$ and $G_\o$. Thus the resulting expression for $\<\Psi^*\Psi\>$
will become gauge dependent. In our simple model this is not really
disturbing, as
the gauge dependence can be reabsorbed in the normalization of the
wave function. Nevertheless, to get insight into similar problems in more
complicated theories,  it is interesting how
a gauge independent measure can be constructed. The introduction
of a Faddeev-Popov determinant, which is the determinant of ($\d$ denotes
the delta function)
\begin{equation} \left( \begin{array}{cc}  \{\pi_\o, G_+\} &
 \{\pi_\o,G_\o\}  \\
  \{\pi_+,G_+\}  &   \{\pi_+,G_\o\}  \end{array} \right)
  = \left( \begin{array}{cc}  -\pi_+ &  0  \\
  0  &   \pi_+  \end{array}
\right) \, \d ,
\label{FP} \end{equation}
will not lead to a satisfactory result.
To our mind this seems to
be correlated to  the fact that the group generated by
$G_\o$, $G_+$ does not allow for an invariant, non-degenerate bilinear form
on its algebra.

To find an invariant measure let us calculate the action of
$G_\o$ and $G_+$ on $\O=[{\rm d}\pi_+] [{\rm d}\pi_\o]$.
We find [cf.\ \re{FP})]
\be \{ \O,\int G_\o \} = \O, \qquad \{ \O,\int G_+ \}= 0. \la{consO} \ee
As this coincides with the transformation of $\pi_+$, it is
obvious that the expression $\prod (1/\pi_+) \,\O$ and thus its dual
$\prod (\pi_+) \delta[\pi_+ -c]\delta[\pi_\o -t]$ is invariant.
Realizing the constraint \re{dQ}) by a further delta functional, we
end up with
\addtocounter{equation}{1}
$$ \< \Psi , \Phi \> \propto \int
[{\rm d}\pi_\o][{\rm d}\pi_+][{\rm d}\pi_-] \:
 \prod_{x^1} [\exp (\pi_\o) \, \pi_+]
\, \d [\pi_+ -c] \, \d [\pi_\o - t] \d [ \6 Q] \:\ti \Psi^* \, \ti \Phi.
 \eqno(\arabic{equation}a) \la{in} $$                                 
Changing the  variables of integration from $\pi_-(x^1)$ to $Q(x^1)$ we find
$$ \< \Psi , \Phi \> = \int {\rm d}Q_0 \: \ti \Psi^*(Q_0) \,
\ti \Phi(Q_0),  \eqno(\arabic{equation}b) $$                          
with the  normalized $\ti \Psi(Q_0) \propto \ti \Psi[\6 Q=0,Q_0]$.
Note that all divergent factors are compensated by
the transformation of the variable of  integration.
We
may now remove the regularisation introduced after \re{par}) and remain
 again with a one dimensional
 quantum mechanical system
described by an ${\cal L}^2(R_+)$ [or ${\cal L}^2(R)$]
with Lebesgue measure.
A solution such as \re{Lio}) could be also  implemented at this stage
when assigning  some
(arbitrary) weight to the point(s) $Q_0 = Q_{deS}$. This  does not seem very
rewarding, though.
To complete the
equivalence with sec.\ 3.1 we have to apply the Dirac observable (18b)
to our wave function \re{wav1}); we indeed find:
\be \< \Psi , \hat P_0 \, \Phi \> =
\int {\rm d}Q_0 \: \ti \Psi^*(Q_0) \: {\hbar \0 i}{{\rm d}\0 {\rm d}Q_0} \,
\ti \Phi(Q_0). \la{12} \ee
The constraint $P_0 >0$ is then implemented  such as in  the
preceding subsection ($P_0 \lra \sqrt{\hat P_0^* \hat P_0}$).

An alternative way to formulate an inner product in the space of
physical wave functions is to first  recognize  that there is a
natural bijective map $\Psi \lra \ti \Psi(Q_0)$ between this space
and  the space of functions over the variable $Q_0$. An
inner product on this space ${\cal L}^2(R_+)$ [or ${\cal
L}^2(R)$]  is  then implicitely defined by the condition that a
basic set of Dirac observables, in our model \,$\hat Q_0$ and
$\hat P_0$, should be hermitian with respect to this inner
product \cite{Les}. Because of $\hat P_0 := (\hbar / i) \, ({\rm
d}/ {\rm d}Q_0)$ [cf.\,\re{sym})] the hermiticity requirement
obviously fixes the measure $\mu$ within the general ansatz
\be \langle \Psi , \Phi \rangle = \int {\rm d}Q_0 \, \mu(Q_0) \,
\ti \Psi^*(Q_0) \, \ti \Phi(Q_0) \ee
to be independent of $Q_0$; so again we end up with (33b).
Whereas this approach to find an inner product is more straightforward
than the first one,
a generalization of it to models (such as general relativity), where a
basic set of Dirac observables is not (yet) known, seems to be
difficult. The approach leading to (33), on the other
hand, is applicable whenever one finds  good gauge conditions.

Choosing $t$ to be $x^1$--independent, it is suggestive to regard
it as an 'intrinsic' time. In analogy to the classical case
one can then denote this time parameter  by $x^0$.
With this interpretation the
second equation of \re{cons}) can be regarded as a kind of Schroedinger
equation [with a time dependent Hamiltonian --- cf.\ (23b)],
and the (time dependent) coordinate transformation from $Q(\pi_a,\pi_\o =t)$
to the variable $Q$ in (33) as  a
shift to a Heisenberg representation of quantum mechanics (cf.\ also sec.\ 4
for more details).
An obvious generalization would be
$ \pi_+=c_+(x^\mu)$, $\pi_\omega =c_\o(x^\mu)$.
Due to our construction of (33) the resulting quantum theories are
independent of the choice of $c_+$ and $c_\o$.
To get  'physical' results we can  e.g.\ calculate
the expectation value of  the
torsion: Plugging the multiplicative operator $\pi^2$ into
(33a) we find
\be \< \pi^2 \>(t) = {1\0 4\g} \, [\<Q_0\> \, \exp(-t) +  (t-1)^2 +1-\L].
\la{tor} \ee
Although we obtained some
nontrivial dynamics by reinterpreting and generalizing the gauge choice
\re{gauge1}) and although we could calculate e.g.\ \re{tor}),
we are not yet ready to determine  $\< g_{11} \>$ etc.,
for having not  fixed the corresponding gauge freedom.
This will be done  in sec.\ 4.


There is also another related approach \cite{Wit} leading to the
correct Hilbert space: Since the constraints are linear in the coordinates,
the momenta are transformed into momenta under the action of the constraints.
Thus we could regard the functionals $\Psi[\pi_\o,\pi_a]$ on the constraint
surface modulo the flow of the constraints as the physical wave functions.
With the general ansatz  \re{ans}) ($\pi_+ \neq 0$) the Lie derivative of
the constraints yield the dependence on $Q(x^1)$ which reduces to the
dependence on its zero mode due to \re{dQ}). The inner product is constructed
as two paragraphs above.

\subsection{Abelianization}

It is well known that any system of first class constraints
allows a formulation, where the constraints are abelian.
A system of canonical coordinates may then be found such that
the abelianized constraints are part of it.
Unfortunately, in general this canonical coordinates are defined
locally only and they are non polynomial in the original coordinates.
Moreover, it is difficult to find them. For these reasons they
are of minor practical use in most systems.
In our system, however, the abelianization will turn out a powerful tool.

Again let us first assume $\pi_+\neq 0$.
We already know the quantity $Q$ to commute with all the
constraints and $\partial Q$ to be a linear combination of the constraints.
It is thus clear that Q will play a crucial role in the abelianization.
As Q is a combination of the momenta, it commutes with all the momenta
and the Poisson bracket with any of the coordinates on the
configuration space
yields a function of the momenta times the delta-function.
So take a configuration space coordinate, devide it by
 the function of the momenta on the right hand side of its
commutator with $Q$ to end up with a canonical conjugate; e.g.\ for $e_1{}^+$
one obtaines in this way:
\be   \{ {1\0 4\g}\, \exp(-\pi_\o ) \,{ e_1{}^-\0\pi_+}(x^1) , \, Q (y^1)\} =
\d(x^1-y^1).
  \la{poi} \ee
There are no obvious  canonical conjugates to the
other constraints.  But a glance at \re{FP}) suggests to reformulate
 $ G_{\o} $ and  $G_+$ by multiplying  them with a factor  $(1/\pi_+)$.
 So we are led to
\addtocounter{equation}{1}
$$ (\ti \o_1,\ti e_1{}^+,Q;\pi_\o,\pi_+,P)
 \eqno(\arabic{equation}a) \la{can} $$                            
with
$$
 \ti \o_1 = { G_+ \0 \pi_+},
\qquad \ti e_1{}^+ = - { G_\o\0\pi_+}  \eqno(\arabic{equation}b)  $$     
$$  P = -{1\04\g}\, \exp(-  \pi_{\o})
\, { e_1{}^-\over \pi_+}.   \eqno(\arabic{equation}c)  $$                
Since $\ti \o_1$ and $P$ are Lorentz invariant and commute with $\pi_+$ they
obviously commute with $\ti e_1{}^+$. Checking finally that also
 $\ti \o_1$ and $P$  commute, we indeed find that
(38a)  forms a complete set of
canonical coordinates (in a region where $\pi_+\neq 0$).

In a region where $\pi_-\neq 0$ we may, up to signs, exchange the
role of '+' and '-' in the above considerations. We thus find
\be (-{G_-\0\pi_-}, {G_\o\0\pi_-}, Q; \pi_\o,\pi_-,
-{1\0 4\g}\, \exp(-\pi_{\o}) \, {e_1{}^+\over \pi_-})
\la{can2} \ee
to form a set of canonical coordinates.

Within our topological framework  it is near at hand to further
Fourier transform $Q(x^1)$ and $P(x^1)$. This then completes the canonical
splitting of our theory into the gauge sector  and the Dirac sector, the
latter   being spanned by
the conjugates
$Q_0 = \int_{S^1} Q(x^1), \,  P_0=\int_{S^1} P(x^1)$.
[Note
that  the zero modes of (38c) and the corresponding variable
in \re{can2}) coincide due to (5b) and \re{Q})].  The
quantization is now obvious.
Any quantization scheme will lead to a system equivalent to that of
 sec.\ 3.1.

\section{Space--Time and Observables}

As the symmetries of a theory of gravity include diffeomorphisms in space and
time, any Dirac observable (i.e.\ any function on the phase space invariant
under the action of the constraints) is space and time independent.  This is
the reason for the lack of any dynamics within the (classical) reduced phase
space or the corresponding quantum system (cf.\ sec.\ 3.1). In order to
reintroduce the notion of space and time into the theory we have to break the
according symmetries. This is most easily done by gauge conditions.  Measurable
quantities are then defined by the requirement to be invariant under the {\em
remaining} symmetries. There is, of course, some arbitrariness in the choice of
gauge conditions. This arbitrariness reflects the fact that different observers
may have different means to measure quantities.

We have already seen an example in section 3.2: $\pi_\omega$ is a function on
the space time manifold.  Under the restrictions  specified in section 2 the
lines where $\pi_\omega$ is constant provide a foliation of space-time. From
\re{l3}) (or also \re{l1})) we find the leaves to be spacelike. This might
encourage us to choose $\pi_\omega$ as a time variable $x^0 :\equiv t$.
Functions on the constraint surface which depend on $t$ and the Dirac
observables alone, like e.g.\ $\pi^2$ [cf.\ \re{Q})],
are invariant under the flow of those
linear combinations of the constraints which leave the gauge condition
\addtocounter{equation}{1}
$$ \pi_\omega-t=0 \eqno(\arabic{equation}a) $$
invariant.  They thus are measurable quantities in this setting and we may
calculate their expectation values etc. for a given quantum state.
In this way we  regain results like \re{tor}).

Let us mention that the choice of a time variable $t$
does not determine its flow $(\6 / \6 t)$:
 A ($t$--dependent) diffeomorphism in the direction of constant time
will leave the choice of time unchanged while varying the flow of time and thus
the Hamiltonian generating it. In our example the condition (40a) implies
\be \{\pi_\omega,H\}=1. \la{tdot} \ee
With \re{Ham}) we find that the values of the Lagrange multipliers $e_0{}^a,
\o_0$ are restricted by
\re{tdot}), but certainly not completely determined.

In order to quantize quantities like e.g.\ the components of the metric, we
have to fix the coordinate system of the observer by further gauge conditions.
The form of the canonical coordinates
(38) suggests the choice
\addtocounter{equation}{-1}
$$ \pi_+ =1, \qquad \6 P=0. \eqno(\arabic{equation}b) $$
\addtocounter{equation}{1}
Due to (40a) and the first eq.\ of (40b), which have been implemented
already within the approach of sec.\ 3.2 for the special case $c = 1$,
the second eq.\ of (40b)  is
equivalent to $\6 e_1{}^- = 0$.  These gauge conditions together with our Dirac
observables uniquely  determine all the quantities
$(\omega_1,e_1{}^a,\pi_\o,\pi_a)$ on the constraint surface $\hat \Gamma$.
A simple algebraic manipulation yields \re{l3}) with $x^0 = t$. (Note that
the choice of good gauge conditions, turning all first class constraints
into second class constraints,  saves one the integration of the flow
of the Hamiltonian; in more complicated systems this can be a decisive
advantage).
Antisymmetrizing these classical relations, we obtain a one parameter
family of hermitian operators on our Hilbert space:
\addtocounter{equation}{1}
$$  \pi_-(t)  = (1/ 4\g)\,[Q_0 \, \exp(-t) + (t-1)^2 +1 -\L]
  \eqno(\arabic{equation}a) $$
$$  e_1{}^-(t) = -4\g \,P_0 \, \exp t
 \eqno(\arabic{equation}b)  $$
$$ e_1{}^+(t) = (1/2)\, [e_1{}^-(t), \pi_-(t)]_+  \eqno(\arabic{equation}c) $$
$$ \o_1(t) =-(1/2)\,[e_1{}^-(t),E(t)]_+  \eqno(\arabic{equation}d) $$
Thus we can now predict mean values, standard
deviations etc.\ for $g_{11}$ and $\omega_1$.

The time evolution of the operator relations (42) is
unitary. To show this let us
first specify the relation between (33a) and (33b). By means of the
second eq.\ \re{dQ}) we can, instead of performing the shift of variables from
$\pi_-(x^1)$ to $Q(x^1)$, also  directly integrate out all the delta functions
within (33a). This yields:
\be
\< \Psi , \Phi \> =  \int {\rm d}\pi_- \:  \Psi_S^*(\pi_-,t) \,
\Phi_S(\pi_-,t)  \la{inS} \ee
with
\be
\Psi_S(\pi_-,t) \equiv \exp({t\0 2}) \, \ti \Psi(Q_0(\pi_-,t)),
\la{PsiS} \ee
the latter function coinciding with $\ti \Psi $ in (33b) and
$Q_0(\pi_-,t)$ being the function at the righthand side of \re{Q}) in the gauge
(40);
$\pi_-$ is the zero mode of $\pi_-(x^1)$, projected out within the
inner product (33). Due to our gauge choice
(40) (with $t \equiv x^0$)
we have been allowed to drop the phase factors in \re{wav1}),
provided we do not consider derivative operators of some higher order  in
$\pi_\o$ and $\pi_+$.  Since the inner product $ \< \Psi , \Phi \> $ does not
depend on $t$ [cf.\ (33b)], there exists a {\em unitary} operator $U(t)$
satisfying
\be \Psi_S(\pi_-,t) = U(t) \Psi_S(\pi_-,0). \la{U} \ee
Because of $\Psi_H(\pi_-) = \Psi_S(\pi_-,0) = \ti \Psi(-Q_0 +{1\0 2\g} - \l)$,
(33b) differs from the usual Heisenberg representation only by the trivial
bijection
\be \pi_- \lra -Q_0 +{1\0 2\g} - \l. \la{bij} \ee
Since (42a) is  the result of the transition from (33a) to (33b)
applied to $\pi_-$, it is obvious that $\pi_-(t)$ is a unitary
evolution in $t$.  (It differs from  the Heisenberg operator
$(\pi_-)_H(t) = U^*(t) \pi_- U(t)$, corresponding to the time
independent operator $\pi_-$ in the effective Schroedinger
picture,  just by this bijection).

In the Dirac-approach of section 3.2 we had to partially fix the gauge  in
order to define an inner product in the space of physical states. It was not
necessary to formulate a gauge condition for the multiplicative constraint
$\partial Q=0$.  Nevertheless,  wanting to obtain unique results for $\<
e_1{}^- (x^1) \>$ etc., this  is necessary.  The corresponding gauge condition
is implemented as an operator condition on the wave functions in this approach:
\be \6 P \, \Psi = {\d\0 \d \6Q} \, \Psi =0. \la{11} \ee
This guarantees that when operators such as $e_1{}^-(x^1)$ act on $\Psi$
 only its (physical) dependence on $Q_0(\pi_-,t)$ contribute to the
result in (33a). Thus effectively the operator $e_1{}^-(x^1)$ acting on
wave functions in the Schroedinger picture can be replaced by its  constant
mode $e_1{}^-$.  The eqs.\ (18b) and
\re{12}) show that the time independent Schroedinger operator $e_1{}^-$ is
transformed into the righthand side of (42b) within the
transition from (33a) to (33b). Since the latter has been shown to be
practically a Heisenberg picture, the unitary evolution of $e_1{}^-(t)$ is also
obvious. Now the unitarity of the remaining two eqs.\ of (42)
is a trivial
consequence: (42c) and (42d) are  the 'Heisenberg evolution' of $(1/ 2)
[e_1{}^-,\pi_-]_+ $ and
$-(1/  2)\,[e_1{}^- ,E_S(t)]_+$, respectively.
The operator $ E_S(t)$ is given by the first eq.\ of \re{Et}); it is
 explicitely time dependent   in the Schroedinger picture.

So, having chosen a gauge (40), there is a natural
Schroedinger picture associated to the operator evolution in the $\ti
\Psi(Q_0)$--representation. The wave functions \re{PsiS}) obey a usual
Schroedinger equation [cf.\ second eq.\ of \re{cons}) in our gauge]
with a hermitian, time dependent Hamiltonian
\be  h_S(t) = -{1\0  2}[e_1{}^- ,E_S(t)]_+. \ee
  With \re{bij}) and the
replacement $ e_1{}^- \lra -4\g P_0$
in the above, we obtain the explicit
form of the evolution operator $U_0(t)$ generating the time dependence of
(42):
\be U_0(t) = T \exp(-{i\0 \hbar} \int_0^t  {2-t'^2\0 2\g} \,P_0 - {1\0 2}
[Q_0,P_0]_+ \, dt') . \la{evol} \ee To find this operator the Dirac approach of
sec.\ 3.2 was very helpful.  Clearly, in order to have a  (non--trivial)
Schroedinger picture it has been necessary to break at least some of the gauge
symmetries of the reparametrization invariant theory.


To calculate the zero components of the connection and the  zweibein, we have
to investigate the flow of time:
with \re{tdot})   and $\{\pi_+,H\}=\{\partial P,H\}=0$ we find
\be
e_0{}^+ = 1 + f \, \pi_-, \quad e_0{}^- =  f(x^0), \quad
\o_0 = - f\,E,           \la{LC+}  \ee
in which $f(t)$ is an arbitrary function of $t$, $\pi_-$ is
determined by (42a), and $E$ by  the expression on the righthand side of
\re{Et}).
It is remarkable that a complete set of
gauge conditions in the $(\omega_1,e_1{}^a,\pi_\o,\pi_a)$-space
$\Gamma$ does not fix the flow of time completely. Irrespective
of how we choose $f$ the classical solutions can be always brought
into the form \re{l3}) (under the assumptions specified
there). There is also another perspective to see this: Since
\re{l3}) is independent of $x^1$ it is invariant under a
coordinate transformation $x^1 \ra x^1 + F(x^0)$, which is the
most general invariance of \re{l3}). This transformation, on the
other hand, induces $e_0{}^+ \ra e_0{}^+ - e_1{}^+ \,
\dot F$, and analoguosly for the other zero components. Starting
from the light cone gauge $e_0{}^+=1,\, e_0{}^-=0,\, \omega_0=0$, this
transformation yields again  \re{LC+}) [with  $f = -e_1{}^- \dot F$],
when restricting it to the cross section $\bar \G$ of the constraints
and the gauge conditions.
 This generalization of the 'light cone gauge'
allows for a strictly timelike flow of time: at least under our
assumptions it is always possible to choose $f$ such that
$g_{00} >0$. It is interesting that through our prescription
leading to \re{LC+}) also the 'Lagrange multipliers' $e_0{}^a,
\o_0$ became {\em operators} in the Hilbert space for any gauge choice
$f \neq 0$. Again different choices of $f$, which may be
operator valued, correspond to different  observers.

The mechanism described above also works in the opposite direction: One may
first choose the values of the Lagrange multipliers. The according flow of time
then restricts but not completely determines the choice of a gauge in the phase
space $\Gamma$.  For instance the light cone gauge leading to \re{l1}) does not
allow for any $x^0$--dependence of $\pi_+$, the $x^1$--dependence of this
function, however, is
still arbitrary. Having chosen  some combination of the constraints as our
classical Hamiltonian $H$ and being able to integrate the flow of it, we may
also introduce the affine parameter of $H$ as an 'extrinsic time' \cite{Tate}
(having factored out the action of the other constraints). This approach seems
straightforward and suggestive,
but it has the drawback that one has to be able to completely
integrate the equations of motion, what usually is not the case; moreover,
any 'extrinsic' time is clearly equivalent to some 'intrinsic'
one introduced by gauge conditions.

It is a special feature of our system that there exists a gauge such that
the $x^1$--dependence of the solutions drops out completely. To obtain
explicitely space--time dependent operators we could  choose a gauge
\be \pi_\o = x^0, \quad \pi_+ = A(x^1), \quad \6 e_1{}^- =0   \la{xgauge} \ee
in which $A(x^1)$ is an arbitrary nonvanishing periodic function,
e.g.\ $A(x^1) = 2 + \sin (x^1/2\pi)$.
Analoguosly to above
the gauge conditions \re{xgauge}) determine uniquely the fields of
$\bar \Gamma$  in terms of the Dirac
observables:
$$  \pi_-(x^1,t)  = - {1\0 A(x^1)}\,[Q_0 \, \exp(-t) - {1\0 4\g} \, (t-1)^2
+ \l - {1\0 4\g}], $$
$$e_1{}^-(t)  = {P_0 \0 \int_{S^1} A} \, \exp t, \qquad
e_1{}^+(x^1,t) ={ e_1{}^-(t)\, \pi_-(x^1,t) \0 A(x^1)},  $$
$$ \o_1(x^1,t) =- {\6 A(x^1) + E(t) \, e_1{}^-(t)\0 A(x^1)}, $$
where $E$ is given by the righthandside of \re{Et}),
and restrict the gauge choice for the zero
components to:
$$ e_0{}^+ = {1 + e_0{}^- \, \pi_- \0 A}, \quad e_0{}^- =
{f(x^0)\0 A}, \quad  \o_0 = - {e_0{}^-\,E\0 A}.            $$
Choosing some gauge for $f(x^0)$ and $A(x^1)$, we could now calculate
$\< g_{\m\n} (x^\m) \>$, $\< \o_{\m} (x^\m) \>$,
$\< \Delta g_{\m\n} (x^\m) \>$, $\< \Delta \o_{\m} (x^\m) \>$, etc.



\section{Conclusion}

We have succeeded in
quantizing the model \re{action}) under
the assumption that the corresponding classical solutions lead to a
space time $M$ of the 'physical' form $M = S^1 \times R^1$
with spacelike $S^1$
and the assumption
$\L \equiv 4\g\l <0$ (in sec.\ 3.2 also under more general assumptions).
The simple structure of the phase space in this restricted model
allowed us to apply different methods of quantization, to
compare the results, and to elucidate conceptual problems of
quantum gravity  like the relation between Dirac observables,
gauge conditions, and measurable quantities --- in a framework, where
the connection between classical and quantum expressions becomes
very clear.


{}From our considerations in chapter 2 we conjecture that the quantization
of models with other values of
$\Lambda$ and with a more general
topology
will come down to the quantization of some finite dimensional
phase space with
a more complicated topology.
In this more general framework the Liouville theory,
which is  the de Sitter solution of our theory, would be
included. Interesting questions like the one of topology
changing could be addressed. We thus think that a detailed
analysis of the general theory would be desireable.

It would be also interesting to compare our results to still further methods
of quantization like e.g. the BRST quantization. In \cite{Jap} the nilpotency
of a quantum version of $c^i G_i + \bar c_i C^i{}_{jk} c^jc^k$ has
been shown. [$G_i =
(G_a, G_\o)$, $C$ are the structure functions, and $c$, $\bar c$ the ghosts
and antighosts, respectively]. The study of the cohomolgy problem of this
operator would be the next step. Another promising area for investigations
seems the  coupling of (1) to matter fields \cite{neu}.


\vspace{5ex}
{\Large\bf Acknowledgements}
\vspace{3ex}

The authors are grateful to W.\ Kummer for drawing their attention
to the model and reading the manuscript and to  F.\ Haider for helping
with the calculations leading to \re{wav1}). Furthermore they
want to thank H.\ Grosse for valuable comments and H.\ Balasin,
M.O.\ Katanaev, P.\ Presnajder, D.\ Schwarz, and H.\ Urbantke for discussions.

\end{document}